\documentclass{article}

\usepackage[utf8]{inputenc}
\usepackage{graphicx}

\usepackage[final]{neurips_2022}

\usepackage[utf8]{inputenc} 
\usepackage[T1]{fontenc}    
\usepackage{hyperref}       
\usepackage{url}            
\usepackage{booktabs}       
\usepackage{amsfonts}       
\usepackage{nicefrac}       
\usepackage{microtype}      
\usepackage{xcolor}         
\title{Learned 1-D advection solver to accelerate air quality modeling}

\author{
  Manho Park \\
  Department of Civil and Environmental Engineering\\
  University of Illinois at Urbana-Champaign\\
  Urbana, IL 61801 \\
  \texttt{manhop2@illinois.edu} \\
  \And
  Zhonghua Zheng \\
  Department of Earth and Environmental Sciences \\
  The University of Manchester \\
  Manchester M13 9PL, United Kingdom \\
  \texttt{zhonghua.zheng@manchester.ac.uk} \\
  \AND
  Nicole Riemer \\
  Department of Atmospheric Sciences \\
  University of Illinois at Urbana-Champaign \\
  Urbana, IL 61801 \\
  \texttt{nriemer@illinois.edu} \\
  \And
  Christopher W. Tessum\thanks{Corresponding author. https://ctessum.cee.illinois.edu}  \\
  Department of Civil and Environmental Engineering\\
  University of Illinois at Urbana-Champaign\\
  Urbana, IL 61801 \\
  \texttt{ctessum@illinois.edu} \\
}

\begin{document}

\maketitle

\begin{abstract}
  Accelerating the numerical integration of partial differential equations by learned surrogate model is a promising area of inquiry in the field of air pollution modeling. Most previous efforts in this field have been made on learned chemical operators though machine-learned fluid dynamics has been a more blooming area in machine learning community. Here we show the first trial on accelerating advection operator in the domain of air quality model using a realistic wind velocity dataset. We designed a convolutional neural network-based solver giving coefficients to integrate the advection equation. We generated a training dataset using a 2nd order Van Leer type scheme with the 10-day east-west components of wind data on 39$^{\circ}$N within North America. The trained model with coarse-graining showed good accuracy overall, but instability occurred in a few cases. Our approach achieved up to 12.5$\times$ acceleration. The learned schemes also showed fair results in generalization tests.
\end{abstract}

\section{Introduction}
Numerical integration of partial differential equations (PDEs) is a core element of air quality model. To run the air quality model one should solve coupled PDEs within many grid boxes and multiple time steps. Since solving many PDEs requires a huge amount of computational cost, the invention of a fast and accurate solver with has been always welcomed. Recent advancement in physics-informed machine learning \citep{karniadakis2021physics, kashinath2021physics} has gained popularity to emulate existing solvers and researchers are seeking a \textit{pareto} optimum between speed and accuracy.

So far, research efforts on learning air quality models for acceleration have mostly focused on chemistry solvers. \cite{kelp2020toward} developed the encoder-operator-decoder structure neural network and achieved $\times$260 speedup in emulation of Carbon Bond Mechanism Z coupled to the Model for Simulating Aerosol Interactions and Chemistry. \cite{huang2022neural} showed their neural integrator could solve two benchmarking problems (the H$_2$O$_2$/OH/HO$_2$ System and the Verwer System) with acceleration in at least one order of magnitude. 

Although this type of research is an emerging area, the potential of a learned solver for transport operators in air quality modeling has not been actively investigated. However, there could be possible speedup by using a learned transport operator since researchers in computational fluid dynamics (CFD) already showed machine-learned acceleration. \cite{kochkov2021machine} and \cite{stachenfeld2021learned} showed convolutional neural network (CNN) based coarse-graining solver could accelerate CFD solvers. \cite{zhuang2021learned}'s learned discretization to estimate coefficients in advection solver achieved 1.8$\times$ faster computing by 4$\times$ coarsening. The more examples of machine-learned fluid dynamics could be found from the review papers by \cite{brunton2020machine} and\cite{kutz2017deep}.

Our study examines the potential of a learned advection operator for computational acceleration of air quality modeling. To build a foundation for future analysis, we first explore the potential of learned solver in 1-D advection. We used a realistic wind velocity instead of a synthetic velocity to generate our baseline dataset. Since the modeling domain in the global air quality model is not rectangular, but spherical, evaluation of the model skills in different grid sizes is critical to generalization. We tested the model's generalization ability in different latitudes which have different grid sizes. Also, we tested the model's ability to integrate the wave from an initial condition shape out of training regime.

\section{Numerical advection}
We simulated a passive scalar advection in a horizontal line passing 39.00$^{\circ}$N of North America (130$^{\circ}$W - 60$^{\circ}$W). The spatial grid size was 0.3125$^{\circ}$. We simulated advection using east-west components of wind data from 1 to 10 January 2019 with 5-minute intervals. We obtained wind data from GEOS-FP of NASA Global Modeling and Assimilation Office \citep{geosfp}. We used a square initial condition with a 10$^{-7}$ on the central 1/3 of domain, while other areas have 0. We implemented the L94 advection scheme \citep{lin1994class} in the Julia computing language \citep{bezanson2012julia}. 

After generating this dataset by numerical integration, we down-sampled those wave datasets in lower resolutions in both space and time. To conserve mass, we averaged the scalar values in down-sampling. The sample resolutions are $\times$1, $\times$2, $\times$4, $\times$8, and $\times$16 in space and $\times$1, $\times$2, $\times$4, $\times$8, $\times$16, $\times$32, $\times$64 in time, so we have 35 different cases from one scenario.

\section{Learned advection using a convolutional neural network}
\begin{figure}[htp]
    \centering
    \includegraphics[width=\textwidth]{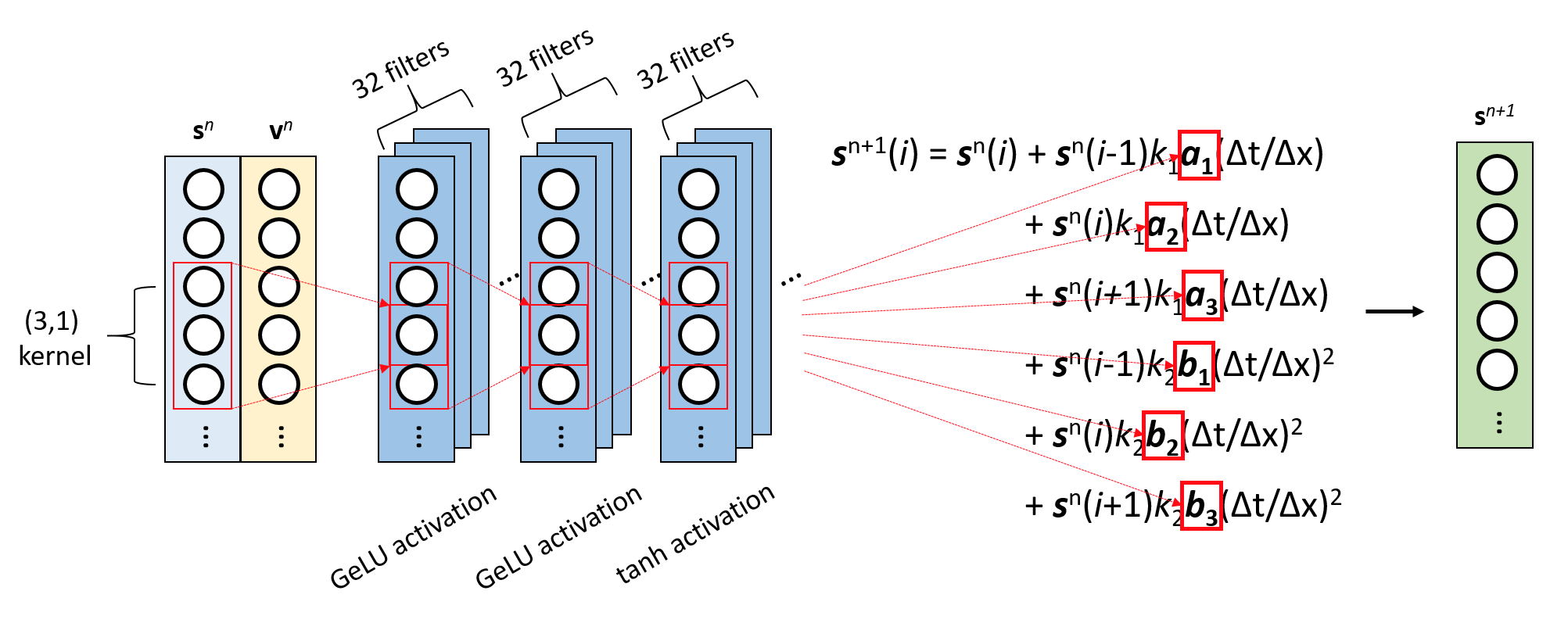}
    \caption{Illustrative diagram of convolutional neural net advection operator}
    \label{fig:Model}
\end{figure}

\textbf{Figure \ref{fig:Model}} illustrates the design of the CNN-based surrogate advection solver. Our surrogate equation has information of $\Delta$t/$\Delta$x, which is critical in earth system modeling domain since grid spacing $\Delta$x can change along with latitude. The three-layer CNN receives scalar and velocity fields at n$^{th}$ time step as inputs and yields six coefficients to construct surrogate numerical equation. We used two GeLU activations \citep{hendrycks2016gaussian} to resolve sharp gradients \citep{kim2021stiff}, and one hypertangent since the temporal gradient can have both positive and negative signs. We used gradient scaling by adopting k$_1$ and k$_1$ to make k$_1(\Delta$t/$\Delta$x) and k$_2(\Delta$t/$\Delta$x)$^2$ be in the order of 10$^0$.

We used mean absolute error (MSE) in 10-time steps as a loss function to reflect the dynamic nature of advection. To prevent error accumulation, we introduced random noise with 4$\times$10$^{-5}$ magnitude of initial scalar intensity. We used the ADAM optimizer \citep{kingma2014adam} with default parameters in Flux.jl \citep{innes2018flux}, except for the learning rate. We used a decaying learning rate to reach the optimum. After training the model, we evaluated the model performance by feeding only the initial condition and velocity field and assessed if the model could integrate the advection process till a given period. We used a single CPU core from an HPE Apollo 6500 system with dual 6248 Cascade Lake CPUs to evaluate computational time.

\begin{figure}[htp]
    \centering
    \includegraphics[width=\textwidth]{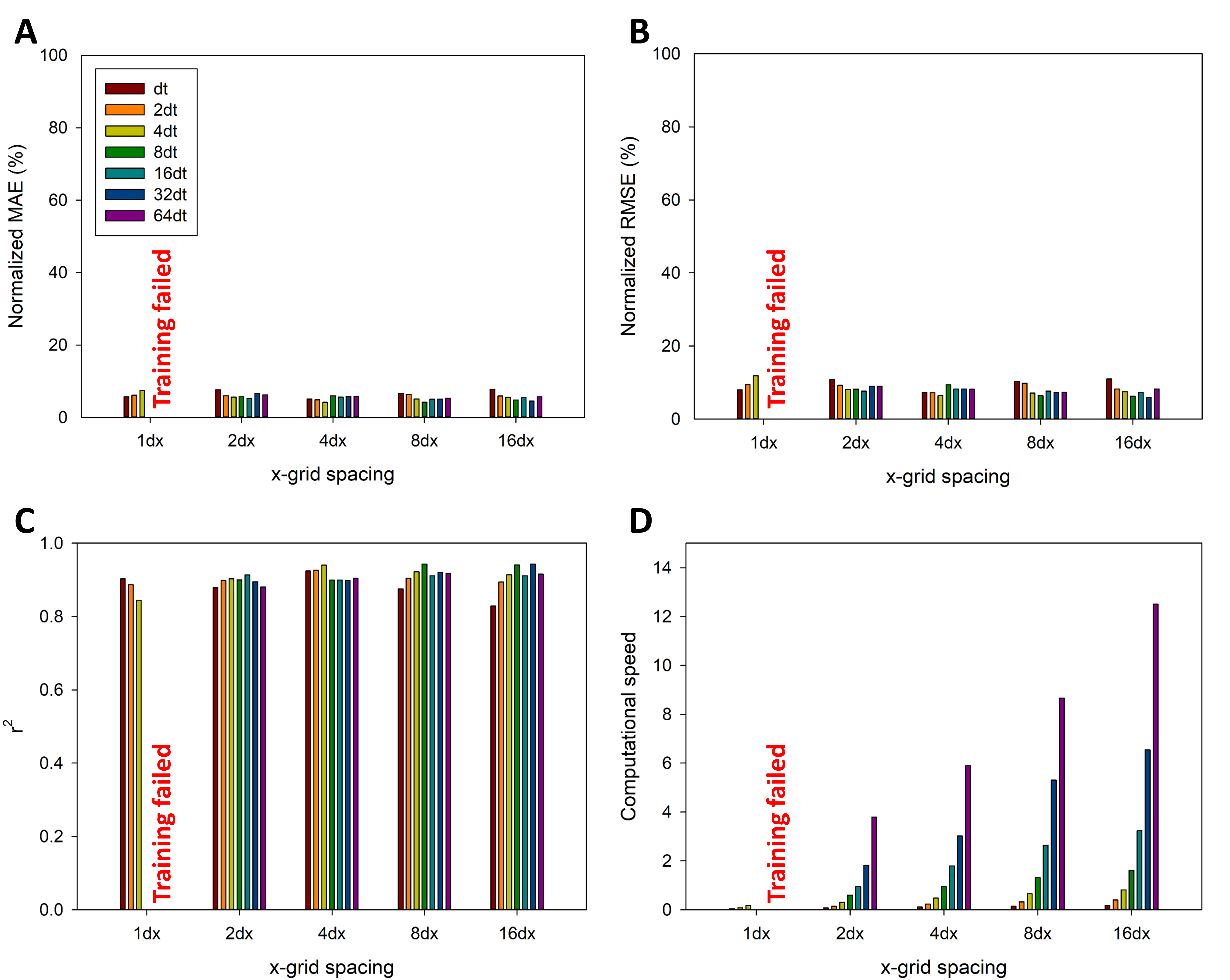}
    \caption{Performance indices of the surrogate learned solver in emulating the training dataset (A: normalized mean absolute error, B: normalized root mean square error, C: r$^2$, and D: speed)}
    \label{fig:Performance}
\end{figure}

\textbf{Figure \ref{fig:Performance}} shows the performance of the learned solver in integrating the training dataset. We normalized MSE and root mean square error (RMSE) by the initial magnitude. Here we should note that the model training was not successful in (1$\Delta$x, 8$\Delta$t), (1$\Delta$x, 16$\Delta$t), (1$\Delta$x, 32$\Delta$t), and (1$\Delta$x, 64$\Delta$t) as the outputs exploded out. Except for those cases, the learned schemes showed fair fidelity since error was usually less than 10 \% (\textbf{Figure \ref{fig:Performance}A} and \textbf{\ref{fig:Performance}B}) and \textit{r}$^2$ was higher than 0.9 (\textbf{Figure \ref{fig:Performance}C}). Maximum acceleration was achieved in (16$\Delta$x, 64$\Delta$t) and this scheme was 12.5$\times$ faster than the original solver (\textbf{Figure \ref{fig:Performance}D}). As seen in \textbf{Figure \ref{fig:Performance}D}, the acceleration in failure cases would be lower than the maximal acceleration case, so we may consider those resolutions to be out of our scope.

\begin{figure}[htp]
    \centering
    \includegraphics[width=\textwidth]{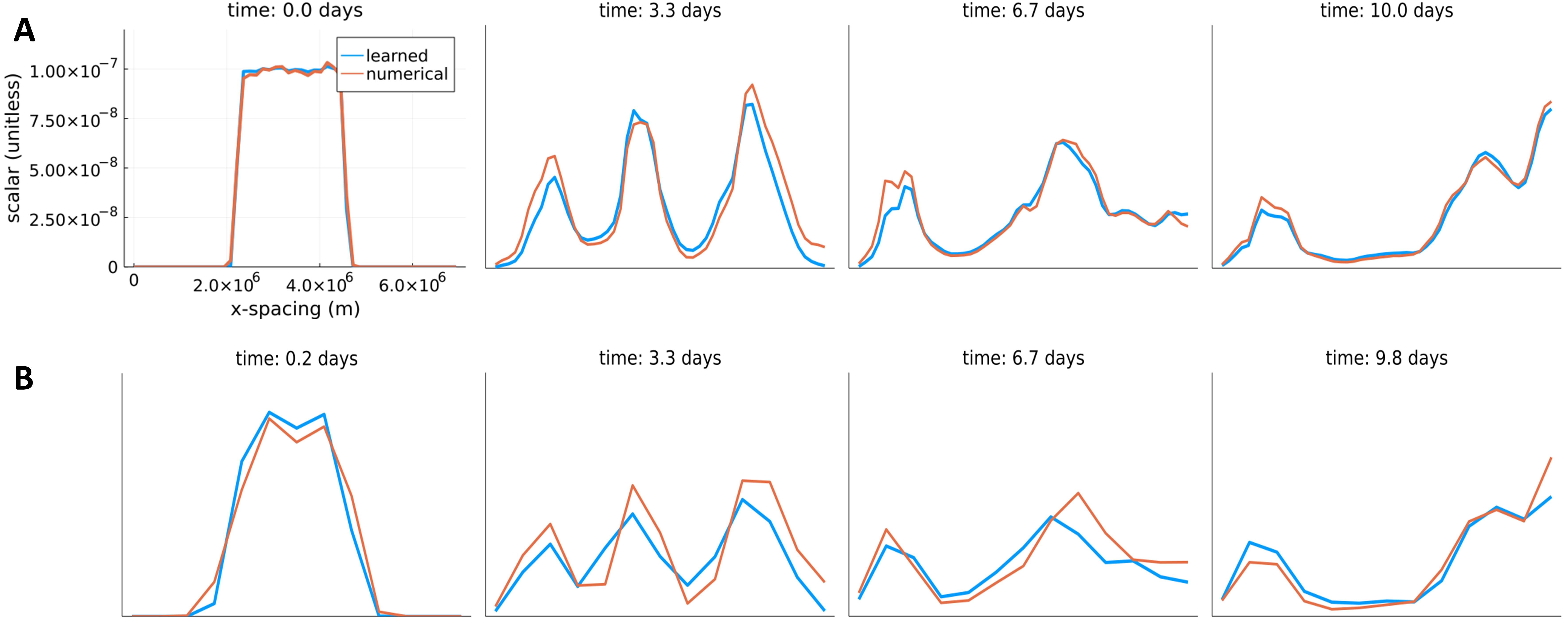}
    \caption{Time series display of both numerical (orange line) and learned (blue line) scalar advection (A: (4$\Delta$x, 4$\Delta$t) and B: (16$\Delta$x, 64$\Delta$t))}
    \label{fig:TimeSeries}
\end{figure}

Time series plots of advection in the resolution with the best accuracy (4$\Delta$x, 4$\Delta$t; \textbf{Figure \ref{fig:TimeSeries}A}) and with the maximum acceleration (16$\Delta$x, 64$\Delta$t; \textbf{Figure \ref{fig:TimeSeries}B}) show the snapshots in the first step from initial condition, 1/3 of time span, 2/3 of time span, and the final step. As seen in \textbf{Figure \ref{fig:TimeSeries}}, we could confirm that the learned model could emulate the coarsened numerical results within fair accuracy.

\section{Generalization}
We tested if our model could integrate the waves from outside of the training regime. We implemented two tests: 1) integrating advection in the Northern area; 2) integrating advection from Gaussian shape initial condition. In the first test, we applied our surrogate solver in a horizontal line passing 45$^{\circ}$N. Though we used 0.3125$^{\circ}$ as grid spacing as we did on the training set, the exact spacing was shorter here because of the earth's spherical shape. In the second test, we fed a Gaussian shape initial condition to be integrated. Any conditions not mentioned are the same as the training.

\begin{figure}[htp]
    \centering
    \includegraphics[width=\textwidth]{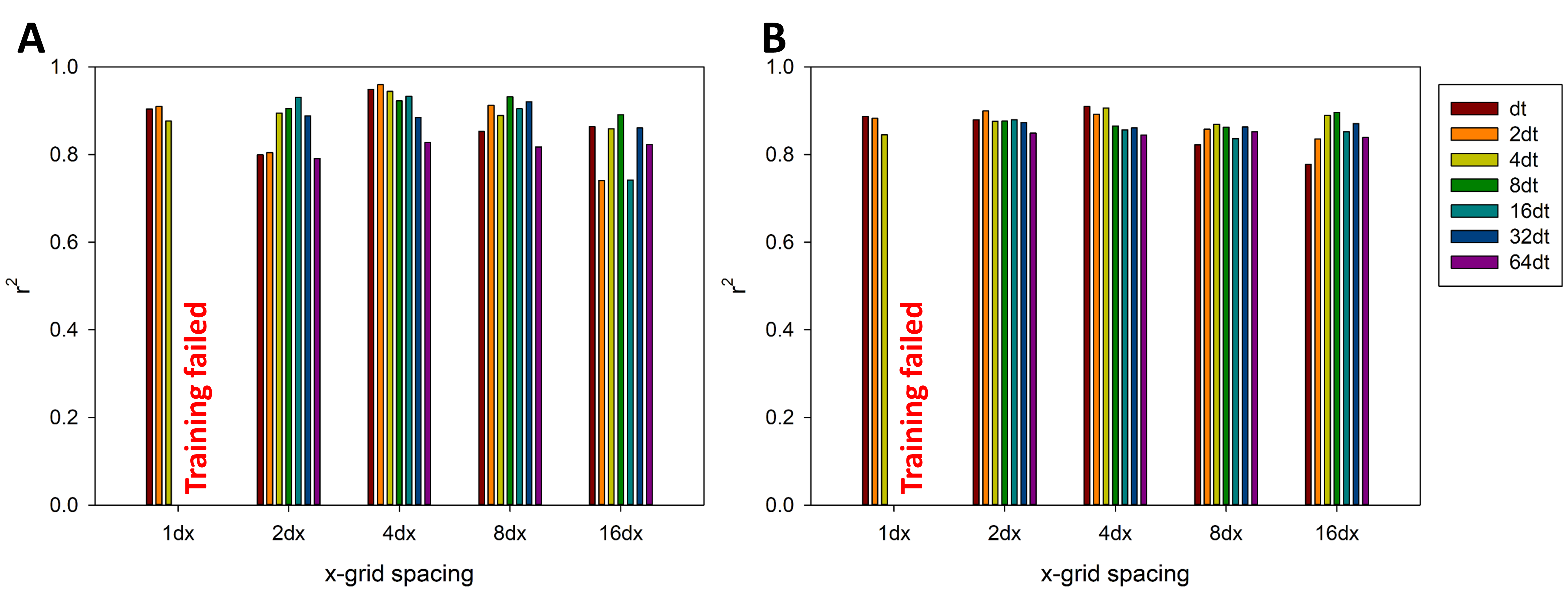}
    \caption{Generalization ability of the learned solver summarized by \textit{r}$^2$ (A: The horizontal line passing 45$^{\circ}$N, and B: Integration with Gaussian shape initial condition)}
    \label{fig:Generalization}
\end{figure}

The results of generalization test are summarized in \textbf{Figure \ref{fig:Generalization}}. \textbf{Figure \ref{fig:Generalization}A} and \textbf{\ref{fig:Generalization}B} show the \textit{r}$^2$ in the horizontal line passing 45$^{\circ}$N and the integration Gaussian shape initial condition, respectively. As seen in \textbf{Figure \ref{fig:Generalization}}, we can use our learned scheme in a regime beyond the training set without substantial performance degradation. We did not test the coarsening cases in (1$\Delta$x, 8$\Delta$t), (1$\Delta$x, 16$\Delta$t), (1$\Delta$x, 32$\Delta$t), and (1$\Delta$x, 64$\Delta$t) because the surrogate models in those cases failed in training.

\section{Limitation}
The most noticeable limitation of this work is that the learned scheme could not work in certain resolutions. One possible explanation is that violations of the CFL condition were most extreme in those cases. Though we increased the stencil size in an attempt to resolve this issue, this approach did not fix the problem. The reason for this instability is an area for future study. Another limitation is that we did not optimize the model training in the light of hyperparameter tuning or rigorous code optimization. We would argue, however, our approach can be promising since it showed robust performance without optimization and there is still room to achieve acceleration. Finally, to use this scheme on the whole globe, we would need more generalization tests on different latitudes to see if the current scheme could work or would need more training on different regimes.

\section{Conclusion}
Our study revealed the learned advection scheme can be used to emulate passive scalar advection in the domain of air quality modeling using real wind-field data. The learned advection solver showed up to $\times$12.5 with fairly accurate integration. The learned solver was robust in generalization even though we only used a single training dataset. This robustness may come from the randomness of realistic wind data, implying choice of a dataset can be crucial in physics-informed machine learning. There could be more potential acceleration by code optimization or high-performance computing. With an appropriate splitting technique, we could extend this learned 1-D advection solver to multi-dimensions and eventually accelerate air quality models.

\section*{Supplementary materials}
The codes and data can be accessed at https://github.com/manozzing/Learned-1-D-advection-solver-with-grid-spacing-physics

\begin{ack}
This work is financially supported by the Early Career Faculty grant of National Aeronautics and Space Administration (grant no. 80NSSC21K1813). MP is supported by the Carver Fellowship and Illinois Distinguished Fellowship.
\end{ack}

\bibliography{citations}

\section*{Checklist}

\begin{enumerate}

\item For all authors...
\begin{enumerate}
  \item Do the main claims made in the abstract and introduction accurately reflect the paper's contributions and scope?
    \answerYes{}
  \item Did you describe the limitations of your work?
    \answerYes{}
  \item Did you discuss any potential negative societal impacts of your work?
    \answerNA{}
  \item Have you read the ethics review guidelines and ensured that your paper conforms to them?
    \answerYes{}
\end{enumerate}

\item If you are including theoretical results...
\begin{enumerate}
  \item Did you state the full set of assumptions of all theoretical results?
    \answerNA{}
        \item Did you include complete proofs of all theoretical results?
    \answerNA{}
\end{enumerate}

\item If you ran experiments...
\begin{enumerate}
  \item Did you include the code, data, and instructions needed to reproduce the main experimental results (either in the supplemental material or as a URL)?
    \answerYes{}
  \item Did you specify all the training details (e.g., data splits, hyperparameters, how they were chosen)?
    \answerYes{}
        \item Did you report error bars (e.g., with respect to the random seed after running experiments multiple times)?
    \answerNA{}
        \item Did you include the total amount of compute and the type of resources used (e.g., type of GPUs, internal cluster, or cloud provider)?
    \answerYes{}
\end{enumerate}

\item If you are using existing assets (e.g., code, data, models) or curating/releasing new assets...
\begin{enumerate}
  \item If your work uses existing assets, did you cite the creators?
    \answerYes{} We cited NASA GMAO webpage to cite GEOS-FP.
  \item Did you mention the license of the assets?
    \answerNA{}
  \item Did you include any new assets either in the supplemental material or as a URL?
    \answerNA{}
  \item Did you discuss whether and how consent was obtained from people whose data you're using/curating?
    \answerNA{}
  \item Did you discuss whether the data you are using/curating contains personally identifiable information or offensive content?
    \answerNA{}
\end{enumerate}

\item If you used crowdsourcing or conducted research with human subjects...
\begin{enumerate}
  \item Did you include the full text of instructions given to participants and screenshots, if applicable?
    \answerNA{}
  \item Did you describe any potential participant risks, with links to Institutional Review Board (IRB) approvals, if applicable?
    \answerNA{}
  \item Did you include the estimated hourly wage paid to participants and the total amount spent on participant compensation?
    \answerNA{}
\end{enumerate}

\end{enumerate}


\end{document}